\documentclass[11pt]{article}%
\usepackage{amsfonts}
\usepackage{amsmath}
\usepackage{amssymb}
\usepackage{graphicx}
\usepackage{float}%
\providecommand{\U}[1]{\protect\rule{.1in}{.1in}}
\newtheorem{theorem}{Theorem}

\newtheorem{lemma}[theorem]{Lemma}

\begin{document}

\title{Construction and separability of nonlinear soliton integrable couplings}

\author{Maciej B\l aszak$^{1,}$\footnote{E-mail: {\tt blaszakm@amu.edu.pl}}, B\l a\.zej M. Szablikowski$^{1, }$\footnote{E-mail: {\tt bszablik@amu.edu.pl}} and Burcu Silindir$^{2, }$\footnote{E-mail: {\tt burcu.yantir@ieu.edu.tr}}\\[3mm]
\small $^1$ Faculty of Physics, Adam Mickiewicz University, Umultowska 85, 61-614 Pozna\'n, Poland\\
\small $^2$ Department of Mathematics, Izmir University of Economics, 35330, Bal\c{c}ova, Izmir, Turkey}

\maketitle

\begin{abstract}
A very natural construction of integrable extensions of soliton systems is presented.
The extension is made on the level of evolution equations
by a modification of the algebra of dynamical fields.
The paper is motivated by recent works of Wen-Xiu Ma et al.  (Comp.
Math. Appl. \textbf{60} (2010) 2601, Appl. Math. Comp.
\textbf{217} (2011) 7238), where new class of soliton systems, being
nonlinear integrable couplings, was introduced.
The general form of solutions of the considered class of coupled systems is described.
Moreover, the decoupling procedure is derived, which is also applicable to several other coupling systems
from the literature.
\end{abstract}

\section{Introduction}

For a given nonlinear integrable dynamical system there usually exists many
different integrable extensions which equations of motion take a triangular
form. They are usually called triangular systems. For Liouville integrable
nonlinear ODE's the interesting class of triangular systems is given in
\cite{K}. For soliton systems we know more examples. Triangular extensions of
the KdV system was considered in \cite{G}. In \cite{Kup} there are constructed linear extensions
of soliton systems, also taking the triangular form,
being by the author called 'dark equations'.  
Another class of linear extensions the so-called linear integrable
couplings of soliton systems was introduced in \cite{FM} and then developed in
\cite{M1} and many other papers. Recently, the theory of nonlinear
integrable couplings of ordinary soliton systems was presented in \cite{M0} and
\cite{M4}.

In the present paper we introduce very natural triangular extensions of basic
integrable nonlinear systems, which from the construction are also integrable.
Moreover their solutions are uniquely determined by the solutions of the basic
equation. This extensions are made on the level of evolution equations
by a modification of the algebra of dynamical fields. The construction is motivated by nonlinear integrable
couplings recently introduced in \cite{M0,M4}. We also propose the decoupling procedure
for the considered class of integrable couplings including those from \cite{M0,M4}
and \cite{Yu,Zh,CZZ}.

In Section~\ref{s2} we introduce the algebra of coupled scalars, which is the underlying algebra
for the nonlinear integrable extensions defined in Section~\ref{s3}.  We
derive soliton integrable couplings, both of field and lattice type. In Section~\ref{s4},
the general form of solution of the coupled systems is obtained. An example of
soliton solutions of the nonlinearly coupled KdV system is presented. In Section~\ref{s5} we
derive a matrix representation of the algebra of coupled scalars and in consequence the matrix
Lax representations for nonlinear couplings constructed in previous sections.
Finally, in Section~\ref{s6}, we prove that any member of the constructed family of coupled systems
separates into copies of the original soliton system. We also show the source of the decoupling procedure.


\section{The algebra of coupled scalars}\label{s2}

Consider $n$-dimensional vector space over field of real numbers $\mathbb{R}$. We define an algebra structure by the following multiplication
\begin{equation}
\mathbf{e}_{i}\cdot \mathbf{e}_{j} := \mathbf{e}_{\max(i,j)},\label{1}%
\end{equation}
were $\mathbf{e}_i$ are the basis vectors.
Let $\mathbf{a}=\sum_{i=1}^{n}a_{i}\mathbf{e}_{i}$, then
\[
\mathbf{a}\cdot\mathbf{b}=
\begin{pmatrix}
a_{1}\\
\vdots\\
a_{n}
\end{pmatrix}
  \cdot
\begin{pmatrix}%
b_{1}\\
\vdots\\
b_{n}%
\end{pmatrix}
 =
\begin{pmatrix}%
c_{1}\\
\vdots\\
c_{n}%
\end{pmatrix}
 =\mathbf{c},
\]
where
\begin{equation*}
c_{i}=a_{i}b_{i}+a_{i}\left(  \sum_{k=1}^{i-1}b_{k}\right)  +\left(
\sum_{k=1}^{i-1}a_{k}\right)  b_{i}.
\end{equation*}
We find that the value of the coefficient $c_{i}$ is given by $a_{i}b_{i}$ plus terms depending on lower order
elements, $a_{k},b_{k}$ with $k<i$. Therefore, we call this algebra as an algebra of coupled scalars.
This algebra is unital, commutative and associative, and $\mathbf{e}_{i}$ are idempotent elements, what follows immediately from the definition  (\ref{1}). The unity element is
$\mathbf{e}_{1}$.

For $n=4$ we have
\[
\begin{pmatrix}%
a_{1}\\
a_{2}\\
a_{3}\\
a_{4}%
\end{pmatrix}
 \cdot
\begin{pmatrix}%
b_{1}\\
b_{2}\\
b_{3}\\
b_{4}%
\end{pmatrix}
  =
\begin{pmatrix}%
a_{1}b_{1}\\
a_{2}b_{2}+a_{2}b_{1}+a_{1}b_{2}\\
a_{3}b_{3}+a_{3}(b_{1}+b_{2})+(a_{1}+a_{2})b_{3}\\
a_{4}b_{4}+a_{4}(b_{1}+b_{2}+b_{3})+(a_{1}+a_{2}+a_{3})b_{4}%
\end{pmatrix}.
\]

In the algebra of coupled scalars, for product of $m$ elements the following formula holds:
\begin{subequations}\label{form}
\begin{equation}
\mathbf{a}^{1}\cdot...\cdot\mathbf{a}^{m}=
\begin{pmatrix}
a_{1}^{1}\\
\vdots\\
a_{n}^{1}%
\end{pmatrix}
 \cdot...\cdot
\begin{pmatrix}%
a_{1}^{m}\\
\vdots\\
a_{n}^{m}%
\end{pmatrix}
  =
\begin{pmatrix}
b_{1}\\
\vdots\\
b_{n}%
\end{pmatrix}
  =\mathbf{b},
\end{equation}
where
\begin{equation}
b_{k}=%
{\displaystyle\prod\limits_{j=1}^{m}}
\left(  \sum_{r=1}^{k}a_{r}^{j}\right)  -%
{\displaystyle\prod\limits_{j=1}^{m}}
\left(  \sum_{r=1}^{k-1}a_{r}^{j}\right)  .
\end{equation}
\end{subequations}

\section{Nonlinear couplings of soliton systems}\label{s3}

Consider a commutative and associative algebra, with respect to the ordinary dot
multiplication, of smooth functions on $\mathbb{R}^{m}$ with $m$ derivations
$\frac{\partial}{\partial x_{i}}:C^{\infty}(\mathbb{R}^{m})\rightarrow
C^{\infty}(\mathbb{R}^{m})$. Let us construct its coupled counterpart
$C_{d}^{\infty}(\mathbb{R}^{m})$, that is an algebra of
functions
\begin{align*}
\mathbf{f}(x) =f_{1}
(x)\mathbf{e}_{1}+...+f_{n}(x)\mathbf{e}_{n} =
\begin{pmatrix}%
f_{1}(x)\\
\vdots\\
f_{n}(x)
\end{pmatrix}
,
\end{align*}
where $x = (x_1,\ldots,x_m)$,
taking values in the algebra of coupled scalars. So, it is
commutative and associative algebra with respect to the multiplication \eqref{1} and
the derivations in $C_{d}^{\infty}(\mathbb{R}^{m})$ can be defines by
$$\frac{\mathbf{\partial}}{\mathbf{\partial}x_{i}}%
:=\frac{\mathbf{\partial}}{\mathbf{\partial}x_{i}}\mathbf{e}_{1}\qquad i=1,...,m.$$
These derivations are well-defined as
\[
\frac{\mathbf{\partial}}{\mathbf{\partial}x_{i}}\mathbf{e}_{1}\cdot(f\cdot
g)=\left(\frac{\mathbf{\partial}}{\mathbf{\partial}x_{i}}\mathbf{e}_{1}\cdot f\right)\cdot
g+f\cdot\left(\frac{\mathbf{\partial}}{\mathbf{\partial}x_{i}}\mathbf{e}_{1}\cdot g\right)
\]
and
\[
\mathbf{f}_{x_{i}} :=
\frac{\mathbf{\partial}}{\mathbf{\partial}x_{i}}\mathbf{e}_{1}\cdot\mathbf{f}=
\begin{pmatrix}%
(f_1)_{x_{i}}\\
\vdots\\
(f_n)_{x_{i}}%
\end{pmatrix}
.
\]

Here we will concentrate on the coupled extension of nonlinear integrable PDE's in $(1+1)$-dimension in
the evolutionary form. Consider an one-field soliton system
\begin{equation}
u_{t}=K[u]\equiv K(u,u_{x},u_{xx,...}). \label{10}%
\end{equation}
Its extension to the system of coupled PDE's takes the
form
\begin{equation}
\mathbf{u}_{t}=K[\mathbf{u}] \equiv K(\mathbf{u},\mathbf{u}_{x},\mathbf{u}_{xx},...),
\label{10a}%
\end{equation}
where
\[
\mathbf{u}=u_{1}\mathbf{e}_{1}+\ldots+u_{n}\mathbf{e}_{n}=
\begin{pmatrix}%
u_{1}\\
\vdots\\
u_{n}%
\end{pmatrix}
 \qquad u_{1}=u,
\]
and in \eqref{10a} the ordinary dot multiplication is replaced by \eqref{1}.

\begin{lemma}
\label{l5 copy(1)}
The system of coupled equations (\ref{10a}) takes the form
\begin{equation}\label{cs}
\begin{split}
(u_1)_{t} &= K\left[u_1\right]\\
(u_{k})_{t} &= K\left[  \sum_{i=1}^{k}u_{i}\right]  - K\left[  \sum_{i=1}
^{k-1}u_{i}\right]\qquad k=2,...,n,
\end{split}
\end{equation}
where $K[\cdot]$ is the same as in \eqref{10}.
\end{lemma}

The proof follows from the power series expansion of $K[u]$ and the relation (\ref{form}).
In fact, it is sufficient to consider homogenous terms as
\begin{align*}
(\mathbf{u}_{i_1x}\cdot\mathbf{u}_{i_2x}\cdot\ldots\cdot\mathbf{u}_{i_mx})_k =
\prod_{s=1}^{m}
\left( \sum_{r=1}^{k}u_{r}\right)_{i_sx}  -\prod_{s=1}^{m}
\left( \sum_{r=1}^{k-1}u_{r}\right)_{i_sx}\qquad m>0,
\end{align*}
where $(\cdot)_k$ means $k$-th coefficient of $\mathbf{u}_{i_1x}\cdot\ldots\cdot\mathbf{u}_{i_mx}$ and $i_s$ are arbitrary nonnegative integers.

Note that the equation \eqref{cs} reconstructs for $u_{1}$ the basic equation. Moreover, for linear PDE's the procedure is trivial as all
related coupled systems are just copies of the basic equation. The extension to the multi-field systems is obvious.

As an instructive example let us consider the $n$-coupled KdV ($n$c-KdV):
\begin{align}
\begin{split}\label{12}
\mathbf{u}_{t}  &  =\mathbf{u}_{xxx}+6\mathbf{u\cdot}\,\mathbf{u}%
_{x}\\
&  \Updownarrow\\
(u_{1})_t  &  = (u_{1})_{xxx}+6u_{1}(u_{1})_{x}\\
(u_{2})_t  &  = (u_{2})_{xxx}+6u_{2}(u_{2})_{x}+6(u_{1}u_{2})_{x}\\
&\  \vdots\\
(u_{n})_t  &  = (u_{n})_{xxx}+6u_{n}(u_{n})_{x}+6\sum_{k=1}^{n-1}(u_{k}u_{n}%
)_{x}.
\end{split}
\end{align}
It is a particular case of the triangular systems. Obviously in dispersionless
limit we get $n$-coupled dispersionless KdV ($n$c-dKdV).

The presented construction is also valid for differential-difference systems. If,
for instance, instead of the derivative $\frac{\partial}{\partial x_{i}}\mathbf{e}_{1}$ one defines on
$C_{d}^{\infty}(\mathbb{R}^{m})$ a shift operator $T_{k}\mathbf{e}_{1}:= \left(  \exp\frac{\partial}{\partial x_{k}}\right)  \mathbf{e}_{1}$, then
\[
T_{k}\mathbf{e}_{1}\cdot\mathbf{f}=
\begin{pmatrix}%
f_{1}(x_{1},...,x_{k}+1,...,x_{m})\\
\vdots\\
f_{n}(x_{1},...,x_{k}+1,...,x_{m})
\end{pmatrix}  .
\]
As an example we consider $n$c-Voltera differential-difference equation:
\begin{align}
\begin{split}\label{13}
\mathbf{v}(x)_{t}  &  =\mathbf{v}(x)\cdot\lbrack\mathbf{v}(x+1)-\mathbf{v}%
(x-1)]\\
&  \Updownarrow\\
v_{1}(x)_{t}  &  =v_{1}(x)[v_{1}(x+1)-v_{1}(x-1)]\\
v_{2}(x)_{t}  &  =v_{2}(x)[v_{2}(x+1)-v_{2}(x-1)]+v_{2}(x)[v_{1}%
(x+1)-v_{1}(x-1)]\\
&  +v_{1}(x)[v_{2}(x+1)-v_{2}(x-1)]\\
&  \vdots\\
v_{n}(x)_{t}  &  =v_{n}(x)[v_{n}(x+1)-v_{n}(x-1)]+v_{n}(x)\left[  \sum
_{i=1}^{n-1}v_{i}(x+1)-\sum_{i=1}^{n-1}v_{i}(x-1)\right] \\
&  +\left(  \sum_{i=1}^{n-1}v_{i}(x)\right)  [v_{n}(x+1)-v_{n}(x-1)].
\end{split}
\end{align}
The two field case, that is $2$c-Voltera, is equivalent to the lattice integrable coupling system from~\cite{M0}.

Each coupled soliton system have a Lax and zero-curvature representation, that can be obtained by replacing an
underlaying algebra by its tensor product with the algebra of coupled scalars.
For the above examples, $\mathbf{L}_t = \left[\mathbf{L},\mathbf{A}\right]$,
where the right-hand side is the usual commutator,
are equations in the algebras of pseudo-differential and shift operators, respectively, over algebra of coupled
functions. Actually, for the $n$c-KdV (\ref{12}) the Lax pair is given by
\begin{equation}\label{lr1}
\mathbf{L}=\partial_{x}^{2}\mathbf{e}_{1}+\mathbf{u},\qquad \mathbf{A}=4\partial_{x}^{3}%
\mathbf{e}_{1}+6\mathbf{u}\cdot\partial_{x}\mathbf{e}_{1}+3\mathbf{u}_{x}
\end{equation}
and for $n$c-Voltera \eqref{13} by
\begin{equation}\label{lr2}
\mathbf{L}=T\mathbf{e}_{1}+\mathbf{v}(x)\cdot T^{-1}\mathbf{e}_{1},\qquad \mathbf{A}=T^{2}%
\mathbf{e}_{1}+\mathbf{v}(x+1)+\mathbf{v}(x).
\end{equation}

As an example of coupled extension of two-field system let us consider $2$-coupled AKNS ($2$c-AKNS), $n=2$:
\begin{align}
\begin{split}\label{akns}
\mathbf{p}_t &= -\frac{1}{2}\mathbf{p}_{xx}+\mathbf{p}\cdot\mathbf{p}\cdot\mathbf{q},\qquad
\mathbf{q}_t = \frac{1}{2}\mathbf{q}_{xx}-\mathbf{p}\cdot\mathbf{q}\cdot\mathbf{q}\\
 &\qquad\qquad\qquad\quad \Updownarrow\\
p_t &= -\frac{1}{2}p_{xx}+p^2q,\qquad
q_t = \frac{1}{2}q_{xx}-pq^2\\
r_t &= -\frac{1}{2}r_{xx}+p^2s+2pqr+2prs+qr^2+r^2s\\
s_t &= \frac{1}{2}s_{xx}-ps^2-2pqs-2qrs-q^2r-rs^2,
\end{split}
\end{align}
where
\begin{align*}
\mathbf{p} = \begin{pmatrix}p\\ r\end{pmatrix},\qquad \mathbf{q} = \begin{pmatrix}q\\ s\end{pmatrix}.
\end{align*}
In fact the system \eqref{akns} was obtained in \cite{Yu} in a different way and is equivalent to the one from~\cite{M4}.
Its zero-curvature representation will be presented in Section~\ref{s5}.

\section{Solutions of coupling systems}\label{s4}

The solutions of the coupled systems (\ref{10a}) are completely determined by
solutions of the basic equations.

\begin{theorem}
\label{t1}Assume that $S^{1},...,S^{n}$ are arbitrary different solutions of the
basic equation $u_{t}=K[u]$. Then, the solution of coupled system (\ref{10a})
is given by
\begin{align}
u_{1}=S^{1},\qquad
u_{k}=S^{k}-S^{k-1},\qquad k=2,\ldots,n. \label{9}%
\end{align}
Moreover, any solution of the coupled system is of the form (\ref{9}).
\end{theorem}

The first part of the theorem is evident, after plugging (\ref{9}) to \eqref{cs} we get
\[
(S^{k}-S^{k-1})_t-(K[S^{k}]-K[S^{k-1}])=0.
\]
The proof of the second part is by induction. If the solution of the basic
equation is $u_{1}=S^{1}$, then the first coupled equation is
\[
(u_{2})_{t}= K[u_{2}+S^{1}]-K[S^{1}]\quad\Longleftrightarrow
\quad(u_{2}+S^{1})_{t}=K[u_{2}+S^{1}],
\]
thus $u_{2}+S^{1}=S^{2}$, where $S^{2}$ is another solution of the
basic equation. The rest of the induction is obvious.
The consequence of the theorem is that for coupled
PDE's any $n$~different solutions of the basic equation build up an appropriate
solution of the coupled system.

\begin{figure}[!ht]
\centering
\includegraphics[scale=1]{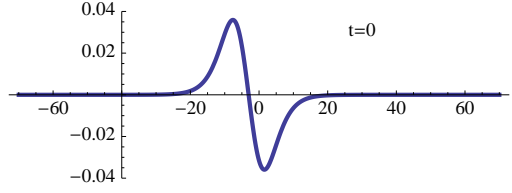}
\caption{Coupled 1-soliton, $u_i = S^i - S^{i-1}$ ($i\neq1$).}
\label{f1}%
\end{figure}

Let us illustrate the whole construction on the
example of solitons of $n$c-KdV \eqref{12} in the Hirota form.
Let us start from $1$-soliton solution. Consider $n$ one-soliton solutions of
the KdV with the same wave velocities and different phases $\gamma_{j}$:
\[
S^{j}=2\partial_{x}^{2}\log(1+f^{j}),\qquad f^{j}=\exp(kx+k^{3}%
t+\gamma_j),\qquad k,\gamma_{j}=const,\quad j=1,...,n.
\]
Then, $1$-soliton solution of (\ref{12}) is given in the form
\[
u_{1}=S^{1}=2\partial_{x}^{2}\log(1+f^{1}),\ \ \ \ u_{j}=S^{j}-S^{j-1}%
=2\partial_{x}^{2}\log\left(  \frac{1+f^{j}}{1+f^{j-1}}\right)
,\ \ \ \ \ j=2,...,n.
\]
Obviously, $u_{1}$ is the ordinary KdV soliton, while remaining $u_{i}$ are
coupled solitons (illustrated in Fig.~\ref{f1}), that differ among themselves by phase arguments.

\begin{figure}[!hb]
\centering
\includegraphics[scale=1]{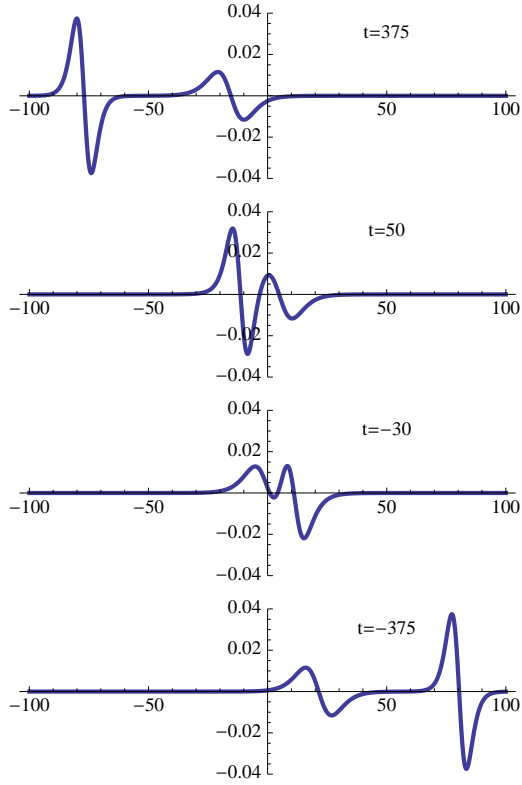}
\caption{Coupled 2-soliton, $u_i = S^i - S^{i-1}$ ($i\neq1$).}%
\label{f2}
\end{figure}

The coupled $2$-soliton solution of $n$c-KdV (\ref{12}) is of the form
\begin{align*}
u_{1}  &  =S^{1}=2\partial_{x}^{2}\log(1+f_{1}^{1}+f_{2}^{1}+af_{1}^{1}%
f_{2}^{1}),\\
u_{j}  &  =S^{j}-S^{j-1}=2\partial_{x}^{2}\log\left(  \frac{1+f_{1}^{j}%
+f_{2}^{j}+af_{1}^{j}f_{2}^{j}}{1+f_{1}^{j-1}+f_{2}^{j-1}+af_{1}^{j-1}%
f_{2}^{j-1}}\right)  ,\ \ \ \ j=2,...,n,
\end{align*}
where
\[
\ f_{1}^{j}=\exp(k_{1}x+k_{1}^{3}t+\gamma_{1j}),\ \ \ f_{2}^{j}=\exp
(k_{2}x+k_{2}^{3}t+\gamma_{2j}),\ \ \ j=1,...,n
\]
and $a=(k_{1}-k_{2})^{2}/(k_{1}+k_{2})^{2}$. Notice that $u_{1}$ is the ordinary
$2$-soliton solution of the KdV, while remaining $u_{i}$ are coupled $2$-soliton
solutions, which interaction is presented in Fig.~\ref{f2}.

\section{Matrix representation of the algebra of coupled scalars}\label{s5}

Consider $n$ quadratic matrices $\mathbf{E}_{k}$ of dimension $n\times n$:
\begin{equation*}
\left(  \mathbf{E}_{k}\right)_{ij}=
\begin{cases}%
1 & \text{if}\quad j=k,\  i\leq k\\
1 & \text{if}\quad j=i,\ i>k\\
0 & \text{otherwise,}%
\end{cases}
\end{equation*}
where $k=1,\ldots,n$.
The matrices $\mathbf{E}_{i}$ constitute
generating elements of commutative and associative sub-algebra of triangular matrices.
Let us call the matrix algebra spanned by $\mathbf{E}_k$ as a pseudo-scalar algebra of matrices, $ps(n)$.

For example if $n=4$:
\begin{align*}
\mathbf{E}_{1}  &  =
\begin{pmatrix}%
1 & 0 & 0 & 0\\
0 & 1 & 0 & 0\\
0 & 0 & 1 & 0\\
0 & 0 & 0 & 1
\end{pmatrix}
\qquad \mathbf{E}_{2}=
\begin{pmatrix}%
0 & 1 & 0 & 0\\
0 & 1 & 0 & 0\\
0 & 0 & 1 & 0\\
0 & 0 & 0 & 1
\end{pmatrix}
\\
\mathbf{E}_{3}  &  =
\begin{pmatrix}%
0 & 0 & 1 & 0\\
0 & 0 & 1 & 0\\
0 & 0 & 1 & 0\\
0 & 0 & 0 & 1
\end{pmatrix}
\qquad \mathbf{E}_{4}=
\begin{pmatrix}%
0 & 0 & 0 & 1\\
0 & 0 & 0 & 1\\
0 & 0 & 0 & 1\\
0 & 0 & 0 & 1
\end{pmatrix}
 .
\end{align*}
Then, a typical element for $n=4
$ is
\begin{align}\label{a}
\mathbf{A}=\sum_{i=1}^{n}a_{i}\mathbf{E}_{i} =
\begin{pmatrix}%
a_{1} & a_{2} & a_{3} & a_{4}\\
0 & a_{1}+a_{2} & a_{3} & a_{4}\\
0 & 0 & a_{1}+a_{2}+a_{3} & a_{4}\\
0 & 0 & 0 & a_{1}+a_{2}+a_{3}+a_{4}%
\end{pmatrix}
.
\end{align}

\begin{lemma}
\label{l1}%
The algebra $ps(n)$ is a matrix representation of the algebra of coupled
scalars defined by the multiplication \eqref{1}.
\end{lemma}

The lemma follows immediately by showing that
\[
\mathbf{E}_{i}\mathbf{E}_{j}=\mathbf{E}_{j}\mathbf{E}_{i}=\mathbf{E}_{\max(i,j)}
\]
for $i,j=1,\ldots,n$.

The Lax representation $\mathbf{L}_t = \left[\mathbf{L},\mathbf{A}\right]$ of
$n$c-KdV \eqref{lr1} and $n$c-Voltera \eqref{lr2} in the matrix algebra
$ps(n)$ are respectively given by
\[
\mathbf{L}=\partial_{x}^{2}\mathbf{E}_{1}+U\mathbf{,\ \ \ \ A}=4\partial_{x}^{3}%
\mathbf{E}_{1}+6U\partial_{x}\mathbf{E}_{1}+3U_{x},
\]
where
\[
U=u_{1}\mathbf{E}_{1}+u_{2}\mathbf{E}_{2}+...+u_{n}\mathbf{E}_{n},%
\]
and
\[
\mathbf{L}=T \mathbf{E}_{1}+V(x)T^{-1}\mathbf{E}_{1},\ \ \ \ \ \ \ \ A=T^{2}\mathbf{E}_{1}+V(x+1)+V(x),
\]
where
\[
V(x)=v_{1}(x)\mathbf{E}_{1}+v_{2}(x)\mathbf{E}_{2}+...+v_{n}(x)\mathbf{E}_{n}.
\]

Taking the tensor product of some Lie algebras, like in the above examples of pseudo-differential and shift operators,
with the pseudo-scalar algebra $ps(n)$ one can derive several examples of
the coupled extensions of known integrable systems.

This is in fact the case of the coupled AKNS systems \eqref{akns}, where its zero-curvature equation,
\begin{align}\label{zc}
\mathbf{L}_t - \mathbf{W}_x +  \left[\mathbf{L},\mathbf{W}\right] = 0,
\end{align}
is from the Lie algebra being tensor product of $ps(2)$ with the loop algebra of $sl(2)$.

The respective generating operators are given by
\begin{align}\label{lax}
\mathbf{L} = \mathbf{E}_{1}\otimes U_{1}+ \mathbf{E}_{2}\otimes U_{2}=
\begin{pmatrix}
U_1 & U_2\\
0 & U_1 + U_2
\end{pmatrix},
\end{align}
where
\[
U_{1}=
\begin{pmatrix}%
-\lambda & p\\
q & \lambda
\end{pmatrix}
,\qquad U_{2}=
\begin{pmatrix}%
0 & r\\
s & 0
\end{pmatrix}
\]
and
\[
\mathbf{W} = \mathbf{E}_{1}\otimes W_{1}+ \mathbf{E}_{2}\otimes W_{2}=
\begin{pmatrix}
W_1 & W_2\\
0 & W_1 + W_2
\end{pmatrix},
\]
where
\begin{align*}
W_1 = \begin{pmatrix} -\lambda^2 + \frac{1}{2}pq & p\lambda - \frac{1}{2}p_x\\
q\lambda + \frac{1}{2}q_x & \lambda^2 - \frac{1}{2}pq\end{pmatrix},\qquad
W_2 = \begin{pmatrix} d & e\\ f & -d \end{pmatrix}
\end{align*}
with
\begin{align*}
d &= \alpha\,\lambda^2 + \frac{1}{2}(1-\alpha)(ps+qr+rs) - \frac{1}{2}\alpha\, pq,\\
e &= ((1-\alpha)r-\alpha p)\lambda + \frac{1}{2}\alpha\, p_x - \frac{1}{2}(1-\alpha)\,r_x,\\
f &= ((1-\alpha)s-\alpha q)\lambda - \frac{1}{2}\alpha\, q_x + \frac{1}{2}(1-\alpha)\,s_x.
\end{align*}
We have left here some freedom related to the parameter $\alpha$
being integration constant from the computation of $\mathbf{W}$.
For various details on derivations of integrable systems we send the reader to \cite{BS} and references therein.
Then from the zero-curvature equation \eqref{zc} we get the following system
\begin{align}
\begin{split}\label{cakns}
p_t &= -\frac{1}{2}p_{xx}+p^2q,\qquad
q_t = \frac{1}{2}q_{xx}-pq^2,\\
r_t &= \frac{1}{2}\alpha\, p_{xx} -\alpha\, p^2q + (1-\alpha)\left (-\frac{1}{2}\,r_{2x} + 2pqr + 2prs + qr^2 + p^2s
+ r^2s \right)\\
s_t &= -\frac{1}{2}\alpha\, q_{xx} +\alpha\, pq^2 + (\alpha-1)\left (-\frac{1}{2}\,s_{2x} + 2pqs + 2qrs + q^2r + ps^2
+ rs^2 \right).
\end{split}
\end{align}
The $2$c-AKNS
\eqref{akns} (also obtained in \cite{Yu}) one obtains for $\alpha=0$. Whereas, the coupled AKNS obtained in \cite{M4}
is generated for $\alpha=-1$. This system is in fact a linear composition of the symmetry \eqref{akns} (the case of $\alpha=0$) and the one obtained from \eqref{cakns} for $\alpha=1$.

In the papers \cite{M0}-\cite{CZZ} there are also other examples of integrable couplings for field and lattice soliton systems
of the same class. All of them are generated by the spectral Lax operators, from the Lie algebra $ps(2)\otimes sl(2)[\![\lambda,\lambda^{-1}]\!]$, being in the form \eqref{lax}, however with different entries of $U_1$ and $U_2$.

\section{Decoupling procedure}\label{s6}

Consider the transformation of the pseudo-scalar algebra $ps(n)$ in the form of a similarity relation
\begin{align*}
	T(\mathbf{A}) :=  S^{-1}\mathbf{A} S,
\end{align*}
where
\begin{align*}
	S_{ij} =
\begin{cases}
1 & \text{for}\quad i\leq j\\
0 & \text{for}\quad i>j
\end{cases}
\end{align*}
and
\begin{align*}
	(S^{-1})_{ij} =
\begin{cases}
1 & \text{for}\quad j=i\\
-1 & \text{for}\quad j=i+1\\
0 & \text{otherwise.}
\end{cases}
\end{align*}

Then, this transformation is apparently an isomorphism of matrix algebras. One finds
that
\begin{align*}
  T(\mathbf{E}_k) = \begin{cases}
1 & \text{for}\quad j=i,\ j\geq k\\
0 & \text{otherwise.}
\end{cases}
\end{align*}

In particular, for $n=4$
\begin{align*}
S = \begin{pmatrix}1 & 1 & 1 & 1\\ 0 & 1 & 1 & 1\\ 0 & 0 & 1 & 1\\ 0 & 0 & 0 & 1\end{pmatrix},\qquad
S^{-1} = \begin{pmatrix}1 & -1 & 0 & 0\\ 0 & 1 & -1 & 0\\ 0 & 0 & 1 & -1\\ 0 & 0 & 0 & 1\end{pmatrix}
\end{align*}
and for $\mathbf{A}$ given by \eqref{a}
\begin{align*}
T(\mathbf{A}): = \begin{pmatrix}a_1 & 0 & 0 & 0\\ 0& a_1+a_2 & 0 & 0\\ 0 & 0 & a_1+a_2+a_3 & 0\\
0 & 0 & 0 & a_1+a_2+a_3+a_4\end{pmatrix}.
\end{align*}

The above transformation naturally extends to any tensor product of some algebra $\mathfrak{g}$
with $ps(n)$, that is $ ps(n)\otimes \mathfrak{g}$.
Let
\begin{align}
\label{l}
  \mathbf{L} = \sum_{k=1}^n \mathbf{E}_k\otimes  A_k,
\end{align}
where $A_k$ belong to $\mathfrak{g}$. Then, the transformation extends to elements of the form \eqref{l}
by the formula
\begin{align}
\label{tl}
T(\mathbf{L}) := \sum_{k=1}^n T(\mathbf{E}_k)\otimes A_k.
\end{align}
Since this transformation is an algebra isomorphism it preserves the Lie algebraic
construction of the coupled systems. Combining \eqref{tl} with the transformation of components:
\begin{align}\label{dc}
\widetilde{A}_k = \sum_{i=1}^k A_i\qquad k=1,\ldots,n,
\end{align}
we decouple the whole construction into $n$ independent copies, that is $ ps(n)\otimes \mathfrak{g}$
decouples into the direct product of $n$ copies of $\mathfrak{g}$.

As consequence of the above decoupling procedure, the linear transformation to new field
variables
\[
\widetilde{u}_k = \sum_{i=1}^k u_i\qquad k=1,\ldots,n,
\]
separates the coupled equations \eqref{10a}-\eqref{cs} to $n$ copies of
the basic equation \eqref{10} in new fields $\widetilde{u}_k$, that is
\begin{equation*}
(\widetilde{u}_k)_{t}=K[\widetilde{u}_k]\qquad k=1,...,n.
\end{equation*}
In particular this is the case of $n$c-KdV \eqref{12} and for $n$c-Voltera \eqref{13} systems.

In the case of the coupled AKNS system \eqref{akns} or \eqref{cakns}
generated by the Lax operator \eqref{lax} the decoupling transformation \eqref{dc} takes the form
\begin{equation*}
\begin{split}
\widetilde{U}_1 &= U_1\\
\widetilde{U}_2 &= U_1 + U_2
\end{split}\qquad\iff\qquad
\begin{split}
\widetilde{p}_1 &= p\qquad \widetilde{q}_1= q\\
\widetilde{p}_2 &= p+r\\
\widetilde{q}_2 &= q+s,
\end{split}
\end{equation*}
where
\begin{equation*}
\widetilde{U}_i = \begin{pmatrix} -\lambda & \widetilde{p}_i\\ \widetilde{q}_i & \lambda \end{pmatrix}\qquad i=1,2.
\end{equation*}
As result the coupled AKNS separates into two copies of the standard AKNS system.

All coupled triangular systems considered in previous sections as well as
these from~\cite{M0}-\cite{CZZ} (as they construction is based on the Lie algebra  $ps(2)\otimes sl(2)[\![\lambda,\lambda^{-1}]\!]$ separate (decouple) into copies of the basic
equations in the same manner. 
In fact the same situation is in the general case, when the underlying Lie algebra has the form
$ ps(n)\otimes \mathfrak{g}$.
This follows from the fact that the above transformation is in the form of similarity relation
which naturally preserves Lie commutator, trace form, Lax and zero-curvature representations,
Hamiltonian structure and other related structures.

\end{document}